\documentclass[a4paper]{article}

\usepackage{INTERSPEECH2020}
\usepackage{subfigure}
\usepackage{tipa}
\usepackage{url}
\usepackage{cite}

\title{Phonological Features for 0-shot Multilingual Speech Synthesis}
\name{Marlene Staib$^1$, Tian Huey Teh$^1$, Alexandra Torresquintero$^1$, Devang S Ram Mohan$^1$,  Lorenzo Foglianti$^1$, Raphael Lenain$^2$\textsuperscript{*}\thanks{\noindent\textsuperscript{*}work done while at Papercup Technologies Ltd.}, Jiameng Gao$^1$}

\address{
  $^1$Papercup Technologies Ltd.,
  $^2$ Novoic}
\email{\{marlene,tian\}@papercup.com}

\begin{document}

\maketitle

\begin{abstract}
Code-switching---the intra-utterance use of multiple languages---is prevalent across the world.  Within text-to-speech (TTS), multilingual models have been found to enable code-switching \cite{zhang2019learning, gutkin2018fonbund, demirsahin2018unified}. By modifying the linguistic input to sequence-to-sequence TTS, we show that code-switching is possible for languages unseen during training, even within monolingual models. We use a small set of phonological features derived from the International Phonetic Alphabet (IPA), such as vowel height and frontness, consonant place and manner. This allows the model topology to stay unchanged for different languages, and enables new, previously unseen feature combinations to be interpreted by the model. We show that this allows us to generate intelligible, code-switched speech in a new language at test time, including the approximation of sounds never seen in training.
\end{abstract}


\section{Introduction}

End-to-end TTS models such as Tacotron 2 are able to generate highly natural-sounding speech, mapping text inputs directly into acoustic outputs \cite{shen2018natural}. Recently, transformations of text used in traditional TTS, such as phonemes \cite{zen2009statistical}, have been found to improve naturalness over characters as inputs within end-to-end models \cite{fong2019comparison}. Similar advantages have been found with phonological features (PFs), such as place and manner of articulation \cite{jakobson1951preliminaries}, when used in addition to, or in place of phonemes as inputs to DNN TTS models \cite{gutkin2018fonbund}. Using phonemes, PFs or both has proven an essential step in training multilingual models \cite{zhang2019learning, gutkin2018fonbund, demirsahin2018unified}. For low-resource settings, performance improvements in both TTS \cite{gutkin2017uniform} and Automatic Speech Recognition \cite{stuker2009porting} were found with PFs over phonemes alone, using either multilingual/multitask or transfer learning from high-resource TTS.

Zhang and colleagues \cite{zhang2019learning} first noted the ability of a multilingual, multi-speaker Tacotron 2 to code-switch, i.e. produce natural-sounding speech of one speaker in two languages, only one of which has been previously seen for that speaker. In their setup, which uses phonemes as inputs, code-switching is only possible for languages within the training data. However, training data is only readily available for a fraction of the world's 5000-7000 languages. The ability to generate appropriate pronunciations, minimally for foreign names, organisations or locations, without requiring large multilingual corpora, is therefore highly desirable in speech applications.

An alternative approach is presented in \cite{li2019bytes},  which explores Unicode bytes as a possible input to a multi-speaker, multilingual Tacotron 2. The advantage over single-valued inputs such as characters or phonemes is that new characters can be added without changing the model topology. This makes it suitable for transfer learning across languages. However, since Unicode only encodes typographic, and not phonological information, nothing is learned in this model about unseen byte combinations, likely requiring at least parts of the model to be relearned entirely when enrolling a new language. 

Gutkin and colleagues \cite{gutkin2018fonbund, demirsahin2018unified, gutkin2017uniform, gutkin2017areal} demonstrate the possibility to synthesise a previously unseen language within multilingual models trained on 9-39 languages, partially with phylogenetic relationships between each other. They use PFs \cite{gutkin2018fonbund, gutkin2017uniform} or a combination of PFs and phonemes \cite{gutkin2018fonbund, demirsahin2018unified, gutkin2017areal} as inputs to their neural, multi-lingual TTS models. They show that various, automatically derived phonological feature sets can be used to either replace or supplement phonemes as input features, yielding improved intelligibility over a phoneme-only baseline across a variety of trained and even untrained languages \cite{gutkin2018fonbund}. To our knowledge, they do not attempt to synthesise any phonemes completely unseen in training. Notably, models which concatenate PFs to phonemes suffer the same constraints on extending the phoneme inventory as phoneme or character-based models, and do not allow for previously unseen phonemes to be synthesised without manual mapping or further training. Finally, these models require a substantial number of training languages to allow for the generation of a new language. 

PFs offer a shared model topology across languages, similar to the ``byte-like'' representation used in \cite{li2019bytes}, and maintain the connection to abstract, phonological categories, while also providing explanatory power on a level closer to the acoustics of an utterance \cite{king2000detection}. While the applicability of a specific PF set to \textit{all} languages is questionable, certain phonological contrasts, such as ``front--back'', have been shown to generalise across various language families \cite{Johny2019CrossLingualCO}. At a lower bound, where phonological categories such as ``fricatives'', ``rounded vowels'', etc. do not share any acoustic properties amongst themselves, PF vectors can be seen as unique identifiers, i.e. the ``byte-version'' of phonemes. If, as we expect, phonological categories have acoustic correlates, we are able to transfer what is learnt to new, unseen or infrequent combinations of sounds. In the case where the acoustics can be disentangled into (somewhat orthogonal) PFs, they would enable us to create new sounds, even including those not present in any human language.

We extend the work in \cite{gutkin2018fonbund, demirsahin2018unified, gutkin2017uniform, gutkin2017areal} by showing that PFs enable code-switching into an untrained language within a small multilingual, or even a monolingual model. Most importantly, we investigate the model's ability to synthesise sounds completely unseen in training (as opposed to an untrained language containing only previously encountered phonemes). While we envision the application of this research to be in code-switching, we conduct our experiments by synthesizing full sentences in an untrained language, marking an extreme case where \textit{all} words are ``code-switched''. Further applications may include TTS for low resource languages, as our experiments simulate a zero-resource setting (irrespective of a particular choice of language).

\section{Phonological features}

\subsection{Feature set}

A range of PF sets have been put forward, including structured primes---nuclear units used to compose phonemes \cite{harris1994english}, binary features---such as ``voice'', ``tense'', ``vocalic'' \cite{chomsky1968sound}, or multi-valued ones---such as place and manner of articulation \cite{jakobson1951preliminaries}. Within the latter, some features can be seen as continuous, for example the horizontal dorsum position for vowels \cite{stuker2009porting}.

In this paper, we use the following set of 10 categorical, multi-valued PFs, 9 of which are directly read from the IPA \cite{ipa1999handbook}: \textit{consonant/vowel, voicing (voiced/unvoiced), vowel frontness, vowel openness, vowel roundedness, stress on vowel, consonant place, consonant manner, and diacritic (e.g., nasalised, velarised)}. Features relating to only vowels are set to \texttt{NULL} for consonants, and vice versa. The tenth feature, \textit{``symbol type''}, is used to integrate symbols that mark, e.g., silences, the end of a sentence, or word boundaries, with all phoneme symbols sharing a single value on this feature. 

Each multi-valued feature is 1-hot encoded into a varying number of binary variables, making up a total of 60 binary features. Phoneme identity is not used as a separate feature, as in other work \cite{demirsahin2018unified, gutkin2017uniform, gutkin2017areal}, since this inhibits the encoding of unseen phonemes in our 0-shot experiments. 

\subsection{Model architecture}

\subsubsection{Baseline  model}
We use a (monolingual/multi-lingual), multi-speaker variant of Tacotron 2 \cite{shen2018natural, zhang2019learning} as our baseline, mapping from phonemes to mel-filterbank features (MFBs). The input text is transformed into phonemes using a linguistic frontend (see \ref{sec:frontend}). The Griffin-Lim vocoder \cite{griffin1984signal} is used to map from MFBs to waveform.

\subsubsection{Phonological features model}
In our proposed PF model, the phoneme embedding table in the baseline is replaced with a single, linear feedforward layer on top of our binary input features. The output dimensionality of this layer is the same as the embedding size of the baseline model. Since the number of binarised PFs in the input feature vector is less than the size of the phoneme inventories used in our experiments (see \ref{sec:exp_data}), the number of parameters used to represent the input is less than in our baseline model.

A phonemic transcription is obtained from text in the same way as in our baseline model, and then further mapped to its PF representation using a dictionary-lookup based on the IPA. Some additional mappings between IPA and resource-specific symbols were necessary, due to the lexical resources used (see \ref{sec:frontend}). Here, we chose the IPA as it is a widely adopted resource. In principle, this framework extends to other PF sets \cite{moran2014phoible, mortensen2016panphon, dediu2016defining} for which automatic extraction is also possible \cite{gutkin2018fonbund}.

\section{0-shot TTS}

\subsection{Synthesizing unseen phonemes}
In our proposed model, our method (AUTO) of synthesizing unseen phonemes is straightforward: PFs of new, out-of-sample (OOS) phonemes are simply inferred from the IPA. PFs can be directly fed to the network without any modifications.

\subsection{Baselines}
To demonstrate the effectiveness of our approach, we consider two baselines of ``0-shot'' TTS in an untrained language: 1) RANDOM, which maps OOS phonemes to a new, randomly initialised (untrained) vector in the phoneme embedding table; and 2) MANUAL, for which we manually map all OOS phonemes to their ``closest'' existing phoneme in the trained embedding table of the Tacotron 2 encoder. We define the closest phoneme as one with a minimal number of differing PFs (e.g., the rounded version of an unrounded vowel), which sounds close to the target sound to a native speaker. In rare cases, a perceptually closer match is found that has less overlap in PF space---e.g., when mapping \textipa{[\;R]} to \textipa{[\*r]} (see \ref{sec:informal_listening} for a discussion). RANDOM serves as a lower bound, demonstrating what can be achieved without additional linguistic knowledge. The comparison of AUTO against MANUAL tests whether our model can go beyond the training data and approximate new sounds from never seen combinations of features, thus outperforming an expert mapping. Minimally, we expect our model to be able to automatically and reliably ``pick`` the closest sound, thereby performing on par with MANUAL.

\section{Experiments} \label{sec:experiment}

\subsection{Data} \label{sec:exp_data}
We use two corpora for experimentation: VCTK \cite{vctk2016}, an open-source, multi-speaker, multi-dialect corpus in English, and Adrianex, a proprietary, multi-speaker corpus in Mexican Spanish. We compare the performance of the baseline and proposed model, using A) VCTK only or B) both corpora combined (MIX). German is used as a target language (see \ref{sec:eval_method}). Statistics, including the number of OOS phonemes in German, are shown in Table \ref{tab:data}. The two data settings are used to explore the relationship between the number of unseen phonemes, and target language intelligibility.
Audio was downsampled to 24 kHz, and 128-dimensional MFBs were extracted every 10 ms over a window of 50 ms.

\begin{table}
  \caption{Data set statistics, including the number of hours in the training set, number of unique phonemes and out-of-sample target phonemes in the test language German (OOS).}
  \label{tab:data}
  \centering
  \begin{tabular}{ l c c c | c }
    \toprule
    \textbf{Corpus} & \textbf{Hours}  & \textbf{Speakers} & \textbf{Phonemes}&  \textbf{OOS}  \\
    \midrule
    VCTK        & $27.5$ & $109$ & $73$  & $14$       \\
    MIX         & $35.4$ & $141$ & $89$ & $9$         \\
    \bottomrule
  \end{tabular}
\end{table}

\begin{figure*}
\centering
\subfigure[baseline model embedding layer]{
\includegraphics[width=.3\textwidth]{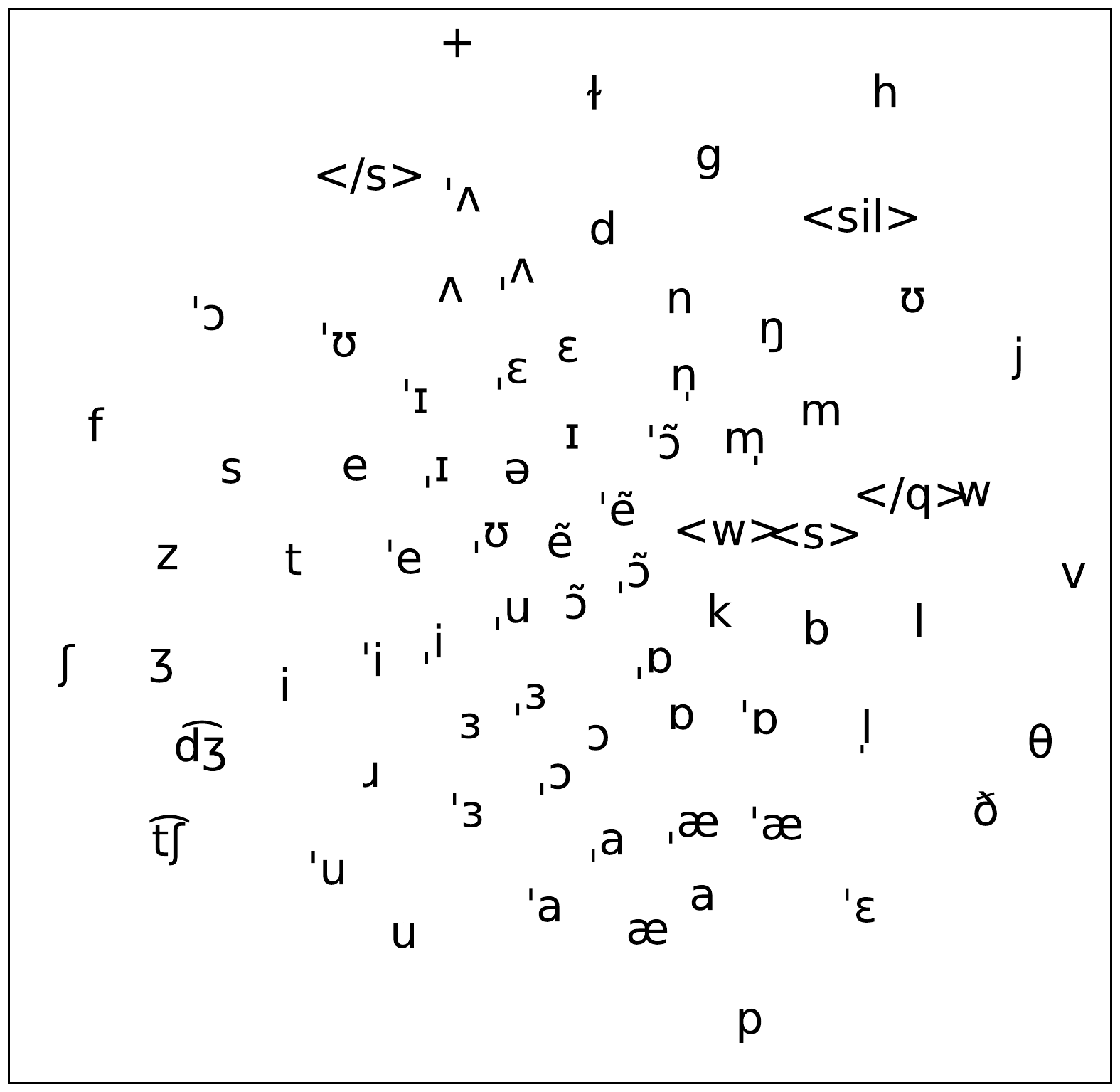}
}
\subfigure[phonological features after first layer]{
\includegraphics[width=.3\textwidth]{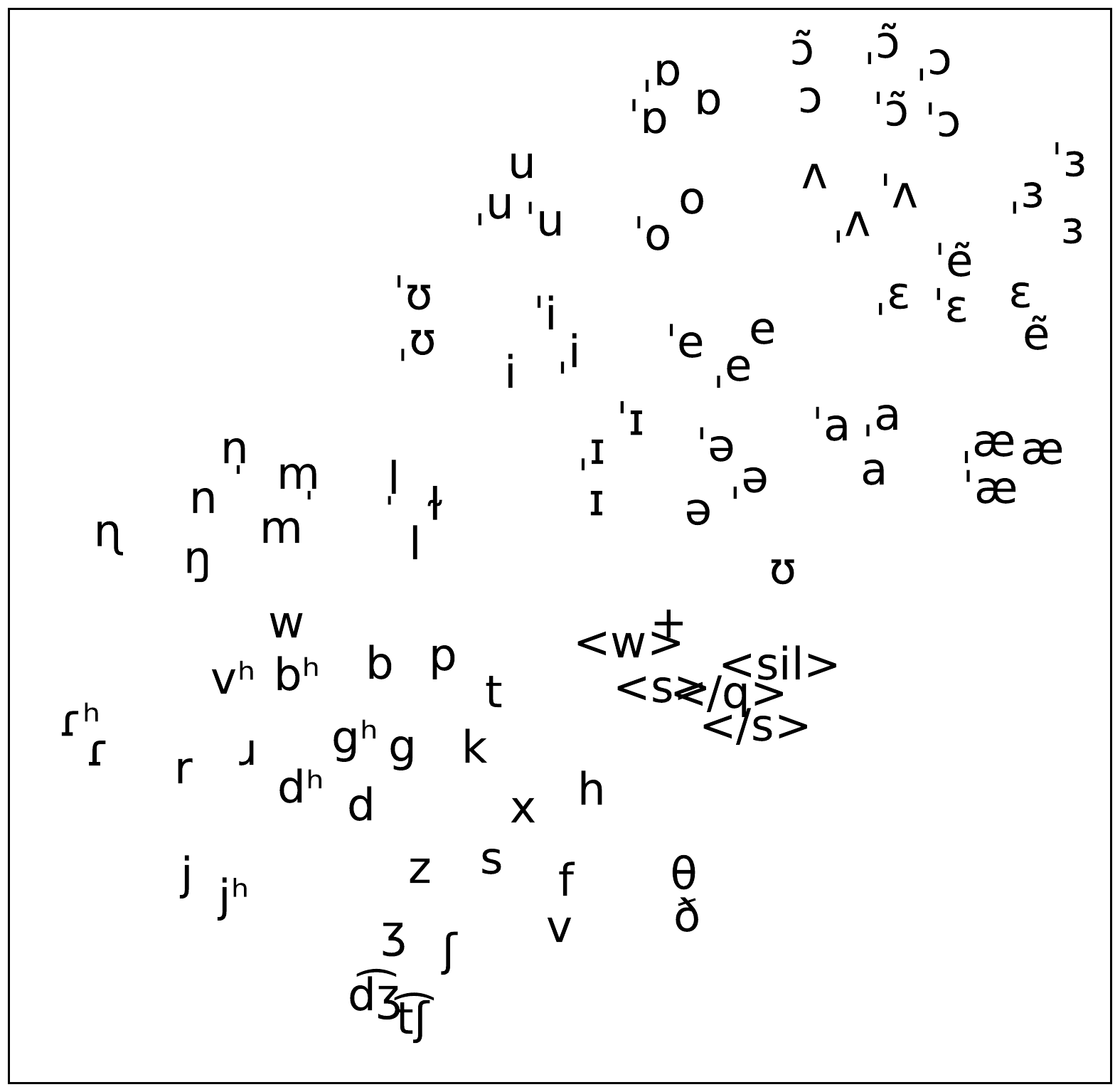}
}
\subfigure[b, with unseen phonemes]{
\includegraphics[width=.3\textwidth]{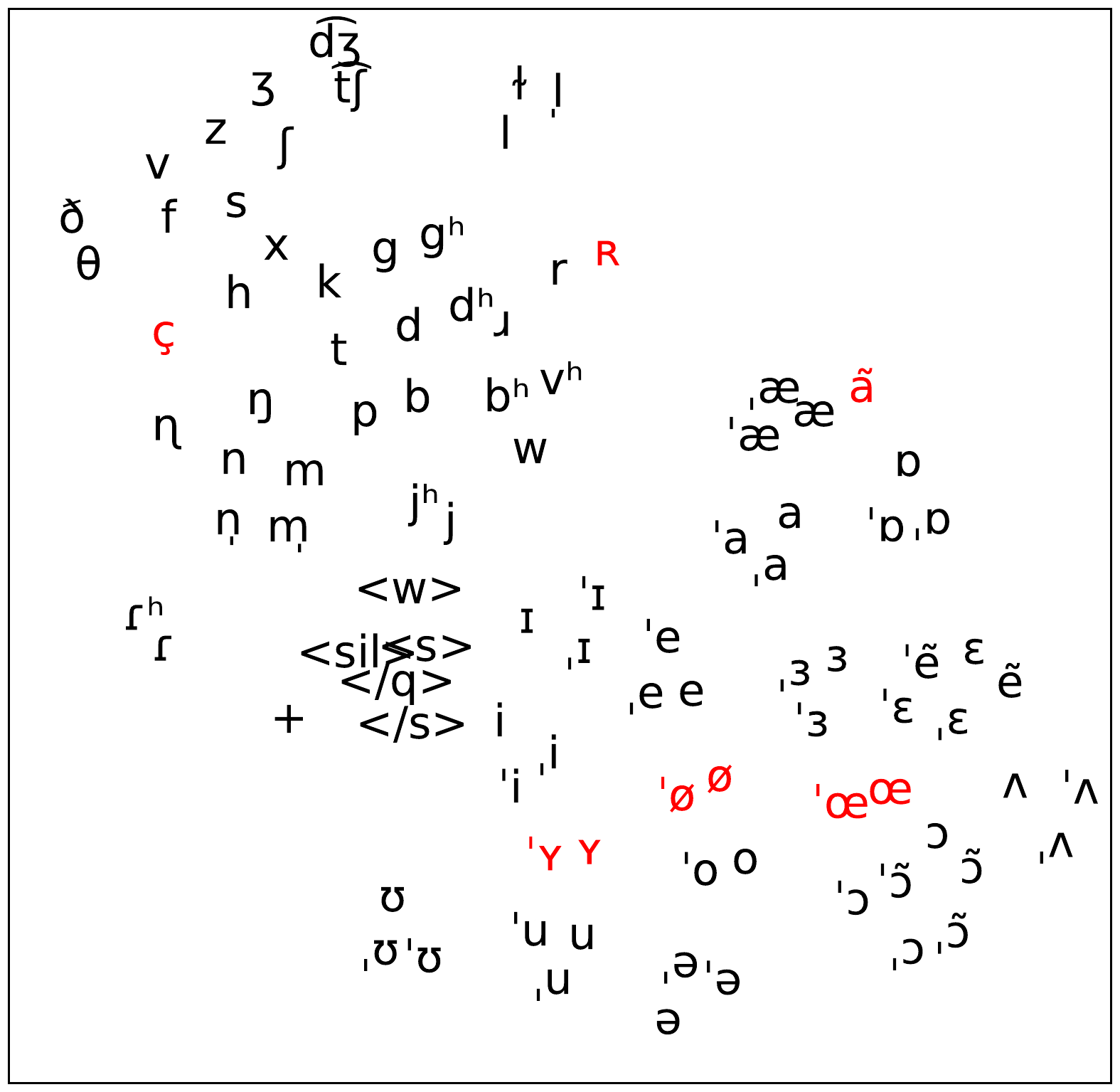}
}

\caption{t-SNE plot of phoneme representations (first encoder layer) from the baseline and proposed model in the MIX data condition. Red in (c): additional phonemes unseen in training. (Note: (c) is a different t-SNE view from (b) of the same representation space.)}
\label{fig:phoneme_embeds}
\end{figure*}

\subsection{Linguistic frontend} \label{sec:frontend}
For English, phonemic transcriptions of the input text are obtained from the Received Pronunciation (RP) version of the Combilex dictionary \cite{richmond2009robust}. For Spanish, pronunciations for each word are obtained using a set of Mexican Spanish pronunciation rules, modified from \cite{oplustil2016multi}. For test sentences in German, the lexicon from the German MARY-TTS voice \cite{MARY2001} is used. For all models including the baseline, resource-specific phoneme-sets need to be mapped into a shared inventory, such as the IPA\footnote{Mapping tables for the listed resources, as well as an IPA-PF lookup dictionary are available at \path{https://github.com/papercup-open-source/phonological-features}}.

Diphthongs, which are represented as a single symbol in the English and German dictionary, are split into their component vowels.
Lexical stress is added to the vowel in a stressed syllable. In the baseline model, stressed vowels are added as a separate symbol in the phoneme embedding table. For diphthongs, stress is added on the first vowel \cite{ladefoged2014course}. Vowel length information, which is available from the German, but not the English lexicon, is discarded.

\subsection{Training protocol} \label{sec:training}
We train our models for 200k iterations, with a batch size of 32 and an initial learning rate of 10e-3, which is decreased to 10e-4 after 100k iterations. Following \cite{shen2018natural}, teacher-forcing in conjunction with dropout is used during training, to align predicted and true MFB output sequences, $\mathbf{\hat{y}}$ and $\mathbf{y}$. The L1 norm between $\mathbf{\hat{y}}$ and $\mathbf{y}$ is used as a loss function for Maximum Likelihood training. To discourage over-reliance on teacher-forcing in the early stages of training, we start with a Prenet layer size of 64, and increase it to 256 after 40k iterations. We found this initial reduction to be essential for the attention network to successfully train across different data settings. All other training parameters are as described in \cite{shen2018natural}.

\subsection{Learned representations in the encoder} \label{sec:repr_space}
We use t-SNE plots \cite{maaten2008visualizing} to inspect the phoneme representation spaces learned by our proposed versus the baseline model (Fig. \ref{fig:phoneme_embeds}). In both models, we observe a meaningful arrangement of phonemes that lends itself to phonological interpretation. For instance, vowels tend to be closer to each other than to consonants, as is the case for voiced versus unvoiced versions of a consonant, and phonemic classes such as nasals, fricatives, stops, etc. However, there are no identifiable clusters in the baseline model, and neighbouring phonemes appear more or less equidistant from each other, likely serving as unique identifiers without further meaningful axes of variation. In contrast, tight clusters emerge in the projection space of our proposed model for vowels versus consonants, nasals, stops, fricatives, etc., suggesting a richer, more meaningful space learnt. Fig. \ref{fig:phoneme_embeds}(b) and \ref{fig:phoneme_embeds}(c) also suggest an intuitive information hierarchy; with PFs clustering by broader group (vowel/consonant/other) first, then---within consonants---by manner second, place third, and everything else (voicing, diacritic) last. 

Moreover, Fig. \ref{fig:phoneme_embeds}(c) shows that new, unseen phonemes get projected into intuitive places within that space without further training. This supports the expectation that, even if the model is unable to interpolate between feature combinations to produce new sounds, it will be able to automatically map unseen sounds to the closest seen sound in feature space, with high accuracy.

\begin{table}
  \caption{Test set statistics, including number of sentences (n sents), sentence length (sent. len.) and Unseen Phoneme Rate (UPR) in percent, in each data setting (where $\mu$ is the mean).}
  \label{tab:testsets}
  \centering
  \begin{tabular}{ l c c c c }
    \toprule
    \multicolumn{1}{l}{\textbf{}} & 
    \multicolumn{1}{c}{\textbf{n sents}} & 
    \multicolumn{1}{c}{\textbf{sent. len.}} &  
    \multicolumn{1}{c}{\textbf{UPR(VCTK)\%}} &  
    \multicolumn{1}{c}{\textbf{UPR(MIX)\%}}  \\
    \midrule
    1        & 30   & 7                 &  $0-5.6$ ($\mu$=$2.9$) & $0-4.3$ ($\mu$=$1.8$)        \\
    2        & 101  & $7-26$  & $0-6.5$ ($\mu$=$2.3$)  & $0-5.4$ ($\mu$=$1.5$)       \\
    \bottomrule
  \end{tabular}
  
\end{table}

\subsection{Formal listening evaluations} \label{sec:evaluations}

\subsubsection{Method} \label{sec:eval_method}

Listening tests were conducted to test whether 1) 0-shot German speech from AUTO is more intelligible than RANDOM, and competitive with MAPPED; 2) AUTO outperforms RANDOM, and is competitive with MAPPED in terms of listener preference; and 3) listener preference is the same for the baseline and the proposed model in the trained language (English). 

For 3, the test set consisted of a random subset of 96 validation set sentences from 6 selected, RP English VCTK speakers (3 male, 3 female). For 1 and 2, sentences were randomly sampled from Wikipedia articles in German, subject to having seven words (or more, as required), all in-vocabulary. The same 6 VCTK speakers were used to synthesise these two test sets. Statistics on the different test sets are shown in Table \ref{tab:testsets}. The Unseen Phoneme Rate (UPR) was calculated as the number of OOS phonemes in a target utterance, divided by the total number of phonemes in that utterance. 

Listener preference was evaluated using A/B preference tests. For intelligibility, listeners were asked to transcribe the generated samples, and word level accuracy was measured. A block design was used to reduce the amount of variability arising from different listeners and different sentence-model pairings. All tests were performed online by native listeners, recruited through Amazon Mechanical Turk \cite{crowston2012amazon}. For preference tests, language proficiency was assessed with a preliminary task of transcribing a sample in the target language. For the transcription task, we discarded responses from participants with an overall Word Error Rate (WER) above 80\%, prior to analysis. Listeners in this category appeared to be non-native speakers or people who had not seriously attempted the task. After filtering, the total number of listeners was 20 for intelligibility and 30 for preference tests.

\subsubsection{Results}

\begin{figure}
\centering
\includegraphics[width=.45\textwidth]{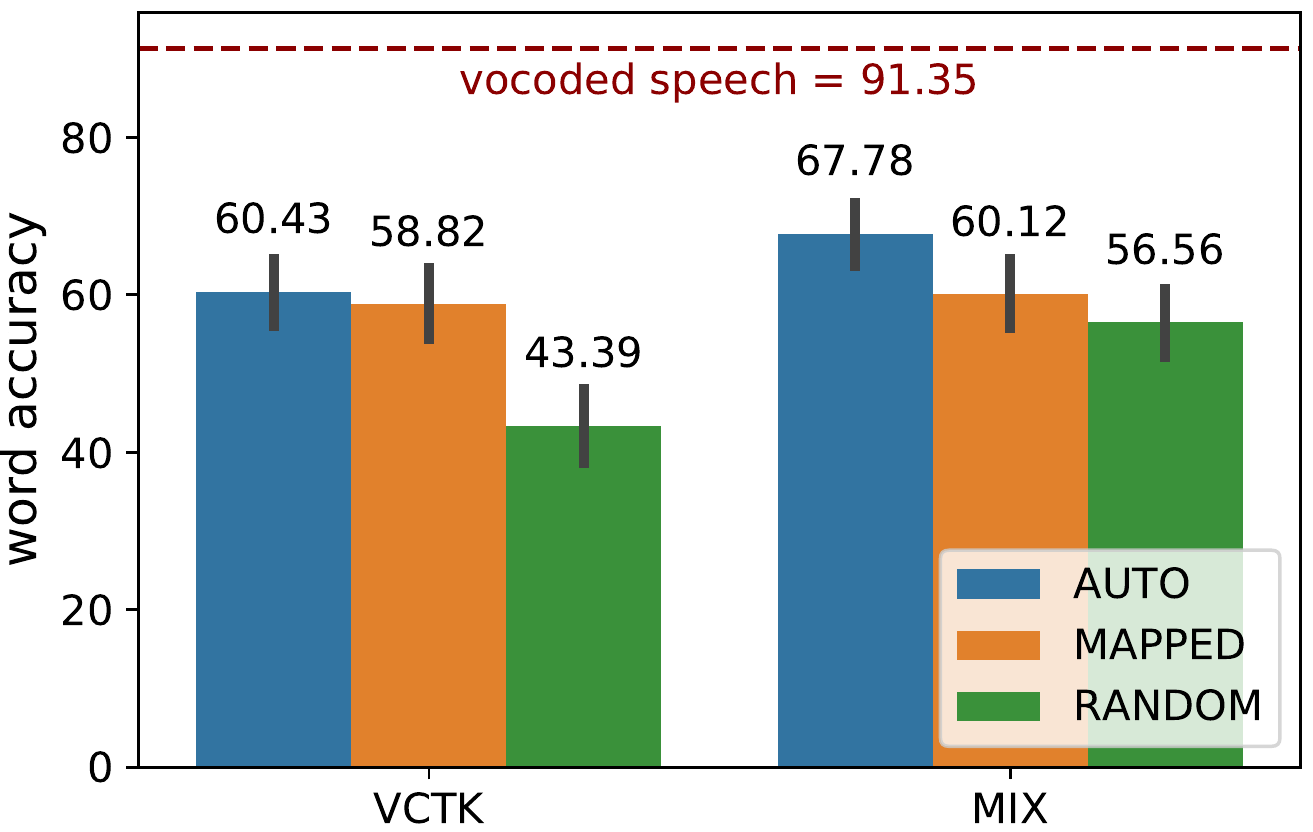}
\caption{Results for intelligibility of 0-shot German.}
\label{fig:intelligibility_results}
\end{figure}

Figure \ref{fig:intelligibility_results} shows that AUTO produces significantly more intelligible speech compared to RANDOM in both data conditions (using pair-wise t-test comparisons; $p < 0.001$), and outperforms MANUAL in the MIX data setup ($p < 0.05$). Interestingly, word level accuracy is similar across data conditions for MANUAL, while AUTO and RANDOM improve with the additional language present in MIX, compared to VCTK alone. The degradation in intelligibility in all tested methods against vocoded human speech is likely due to the strong English accent present in the 0-shot samples. Producing accent-free, 0-shot speech, perhaps relying on more sophisticated representations of typology and language \cite{gutkin2017areal, tsvetkov2016polyglot} or with adversarial losses \cite{zhang2019learning} remains a challenging task for future research.

We find a weak negative correlation between UPR and word accuracy across data settings in RANDOM ($r=-0.12, p<0.05$), but not in AUTO or MANUAL. This suggests that, while RANDOM is exposed to pronunciation (and subsequent perception) errors arising from unseen phonemes, AUTO and MANUAL are both effective strategies to overcome or at least attenuate them. Further research is needed to determine how and why pronunciations in AUTO improve with more languages, more data or a larger input phoneme inventory. Since we were unable to measure a clear relationship between UPR and word accuracy in AUTO, phoneme coverage alone may not fully explain this effect. Another possibility is that, rather than just exploiting a greater number of learned phonemes, AUTO is actually able to learn a more sophisticated phoneme projection space when trained on MIX.

\begin{figure}
\centering
\subfigure[preference for the 0-shot systems]{
\includegraphics[width=.45\textwidth]{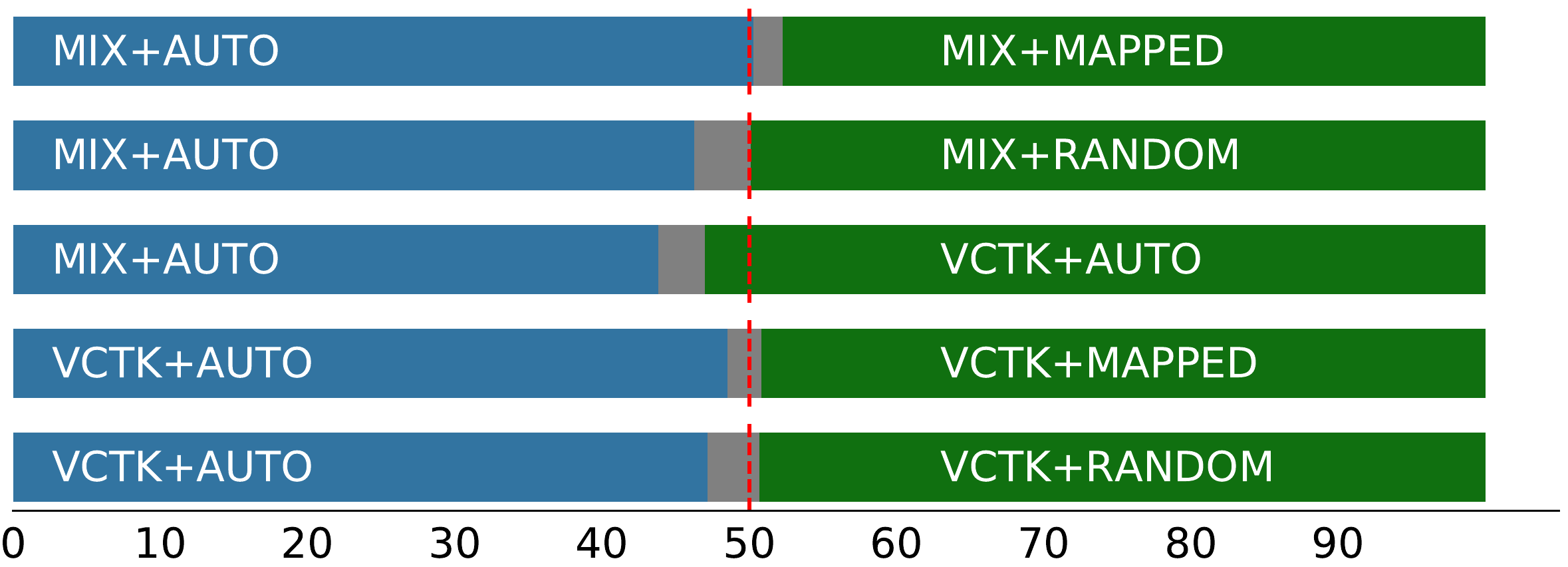}
}

\subfigure[preference in English]{
\includegraphics[width=.45\textwidth]{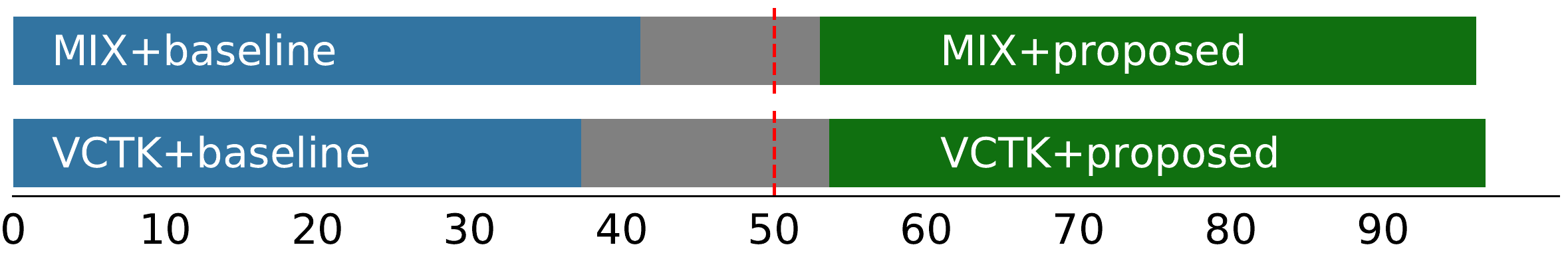}
}
\caption{Results of the listening evaluations.}
\label{fig:preference_results}
\end{figure}

No clear trend in listener preference was found for any of the compared methods (Fig. \ref{fig:preference_results}(a)).
A potential explanation is that listeners' preference was heavily influenced by the strong English accent present in the 0-shot speech, which may have overshadowed the preference for any particular model. It is also conceivable that the improved phoneme representation primarily affects \textit{pronunciation} quality, which manifests itself mostly in intelligibility, over other metrics.

We also did not observe a significant difference in listener preference for the baseline versus our proposed model in English, in either of the data settings (Fig. \ref{fig:preference_results}(b)), indicating that the ability to do 0-shot TTS in a new language in our proposed model does not come at the cost of reduced naturalness or quality in the trained language.

\subsection{Informal listening evaluations}\label{sec:informal_listening}
Through informal listening, we find that most of the OOS phonemes collapse to neighbouring in-sample phonemes in the audio output. Most often, this produces a perceptually agreeable mapping, e.g. from \textipa{[ç]} to \textipa{[S]} and \textipa{[Y]} to \textipa{[I]}. In one case, within the monolingual (VCTK) model, the collapse is inappropriate, mapping the trill \textipa{[\;R]} to the stop [g]. This does not happen for the multilingual (MIX) model, where \textipa{[\;R]} sounds either like the alveolar trill \textipa{[r]}, or approximant \textipa{[\*r]} in the generated output\footnote{Samples can be found at \path{https://research.papercup.com/samples/phonological-features-for-0-shot}}. 
When a feature is completely unobserved in training (such as "trill" in the VCTK data setting), a manual mapping may be preferable to human listeners, e.g., from \textipa{[\;R]} to \textipa{[\*r]} (sharing few PFs, but having a common graphemic symbol 'r').

\section{Conclusion and Future Work}

By replacing the character input in Tacotron 2 with a relatively small set of IPA-inspired features, we were able to create a model topology which is language independent, and allows for the automatic approximation of sounds unseen in training, exceeding or matching the performance of various baselines including a resource-intensive expert mapping approach.

Further research is needed to validate this work for different pairs of source and target languages. This involves a more thorough investigation of the effect of the number of training languages, the amount of overlap between phoneme and PF inventories, as well as other typological and phylogenetic relationships. Another avenue of interest would be to further disentangle PFs, as well as accent, in order to generate unseen combinations of features more accurately. 

\section{Acknowledgements}
We would like to thank our advisor, Simon King, for his input to this research, Doniyor Ulmasov, for his questions which inspired the idea for this paper, and the anonymous reviewers for their valuable feedback.

\bibliographystyle{IEEEtran}
\bibliography{bibliography}

\end{document}